\documentclass[runningheads,a4paper]{llncs}

\usepackage{amssymb}
\usepackage[utf8]{inputenc}
\usepackage[english]{babel}
\usepackage[frozencache,cachedir=.]{minted}
\usepackage{graphicx}
\usepackage{url}
\usepackage{color}
\usepackage{silence}
\usepackage{booktabs}
\usepackage[title]{appendix}
\begin{document}

\mainmatter  

\title{Block the blocker: Studying the effects of Anti Ad-blocking}

\titlerunning{Studying the effects of Anti Ad-blocking}

\author{Rohit Gupta, Rohit Panda}
\institute{Technical University of Munich, Germany\\
\email{rohit.gupta@tum.de, rohit.panda@tum.de}
}

%
%

\toctitle{Lecture Notes in Computer Science}
\tocauthor{Authors' Instructions}
\maketitle

\begin{abstract}
Advertisements generate huge chunks of revenues for websites and online businesses. Ad-blocker and tracker blocking programs have gained momentum in the last few years with massive debates raging on privacy concerns and improving user experience online. Acceptable Ads programme and Anti Ad-blockers are primary elements emerging in recent years that combat ad-blockers.

In this paper, we discuss at length data collection of top websites in the world, Germany, DACH region and news category. We generate feature based A/B testing metrics and employ classifier evaluations on them along with then analysing the result. Our paper also discusses how Anti Ad-blockers impact the economic, legal and ethical usage in Germany alongwith the recent changes in GDPR while taking a look at Acceptable ads programme and Whitelisting.

\keywords{Advertisements, Ad-blocker, Privacy, Anti Ad-blocker}
\end{abstract}

\section{Introduction}
\label{section:intro}

\subsection{Motivation}
Advertisement is a technique used by product and service companies that detail the specific selling points of their products by means of which it becomes known to the masses. Marketers usually place ads at calculated locations for their products such that they are able to draw the attention of potential consumers and persuade them to buy it.

Online advertising or online marketing is a technique of delivering promotional messages to various consumers. It is what drives the economy of most of the World Wide Web businesses. Most modern websites, in general, tend to monetize their user visits. They include certain spaces across their websites aimed at advertisers to come and put their promotional content. An unsaid agreement exists between the website owner and the advertisers on displaying only genuine promotional content and not include malwares or even involve scamming a user.

Ad-blockers have emerged as tools that blocks such advertisements to improve users' web-browsing experience, maintaining privacy, and recently protecting themselves against malware. This has impacted businesses that rely on revenues from advertisements.

Anti Ad-blockers have emerged as an increasingly popular solution. Anti Ad-blockers detect if Ad-blockers are running and use several techniques such as simply notifying the user that the tool interferes with the content and the user-experience or other times when the message blocks the user from accessing the content until they have turned off the Ad-blocker. In extreme cases the goal is to circumvent the tool completely.

Interactive Advertising Bureau (IAB) have been actively working on this topic and have also made a script to "DEAL (Detect, Explain, Ask, Limit)" with such ad-blockers public. They call this "Ad Block Detection Code Access Request" and their source code is available on Github \cite{IAB2017}: 

\textit{https://github.com/InteractiveAdvertisingBureau/AdBlockDetection}

\subsection{Roadmap}
In our paper we propose to study approaches to:
\begin{itemize}
	\item Study the usage of these Anti Ad-blocker scripts and their mechanism.
    \item Collect and analyse top websites in the world using Alexa Website Rankings.
	\item Economic impact of Anti Ad-blockers.
	\item Legality and ethics of Anti Ad-blocking.
    \item Impact of GDPR on Anti Ad-blocking.
	\item Alternatives to Anti Ad-blocking such as Whitelisting and Acceptable Ads programme and also taking a look at Anti Ad-block killers.
    
\end{itemize}

This paper is structured as follows:

\textit{Section 2:} In this section, we discuss the background related to how Anti Ad-blockers work and also take a look at the related work on this topic.

\textit{Section 3:} In this section, we discuss how we collect top websites from different sources. 

\textit{Section 4:} In this section, we discuss how we process the data that we collected including methodology used for our results analysis.

\textit{Section 5:} In this section, we discuss the results for the methodology used in the previous section.

\textit{Section 6:} In this section, we analyze the impact of Anti Ad-blockers such as the economic, legality, ethical aspects including how recent changes in GDPR affects Anti Ad-blockers.

\textit{Section 7:} In this section, we discuss alternatives to Anti Ad-blocking such as Acceptable Ads and Whitelists and also take a look briefly at Anti Ad-block killers.

\textit{Section 8:} We conclude our discussion of our work and provide the roadmap ahead.

\section{Background and Related Work}

\subsection{Background}
Online advertisements or online marketing involves putting out promotional content on the Internet for users. Largely, the economy on the World Wide Web hangs on the fate of online advertisements. Advertisements are put out at different locations on websites such that users are enticed to look at them and react if they are interested in such promotions.

This has led to a stark increase in security and personal concerns and issues. Users argue that they do not want such annoying advertisements on websites because it hampers their area of viewing actual content seamlessly and without obstructions. Some users also say that they could be used to deploy malwares and scams on the web.

Ad-blocker tools are countering this problem by introducing a technique of removing such undesirable advertisements. Some of the famous Ad-blocker tools include Adblock Plus or Ghostery. As of July 2018, more than 10 million users use Adblock Plus on Google Chrome \cite{Google2018} which shows that users are actively seeking to filter out or limit such advertisement promotions.

In a measurement study "Annoyed Users: Ads and Ad-Block Usage in the Wild" by Pujol et. al. \cite{Pujol2015}, they observe that 22\% of active users use Adblock Plus on their browsers.

\begin{figure}
\centering
\includegraphics[height=6.2cm]{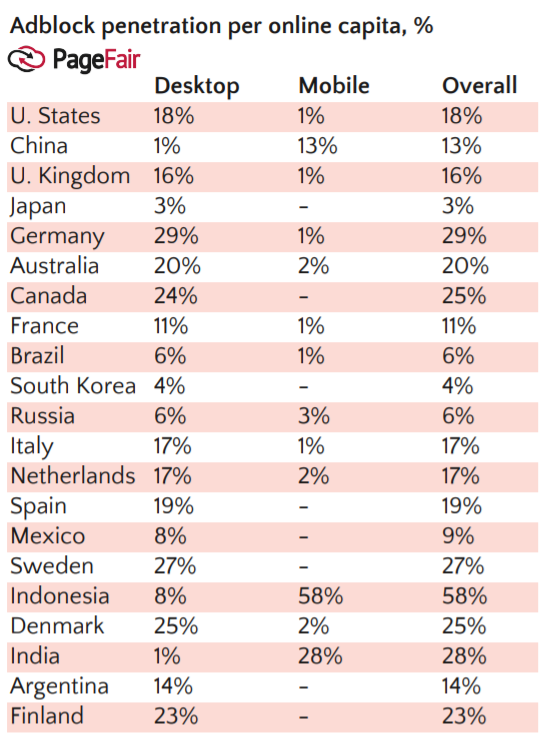}
\caption{Top Ad Markets (ad spend) statistics taken from a PageFair report titled "2017 Global Adblock Report".}
\label{fig:popGermany}
\end{figure}

\textbf{The popularity of Ad-blockers in Germany.} According to a "2017 Global Adblock Report" published by PageFair \cite{popularity2017}, Germany ranked highest in Ad-block installations when it came to Ad-blockers penetration data for desktop (Figure \ref{fig:popGermany}). 

These statistics indicate a tremendous potential for European markets since there is no clear single solution for mobile Ad-block usage. 

\begin{figure}
\centering
\includegraphics[height=7.2cm]{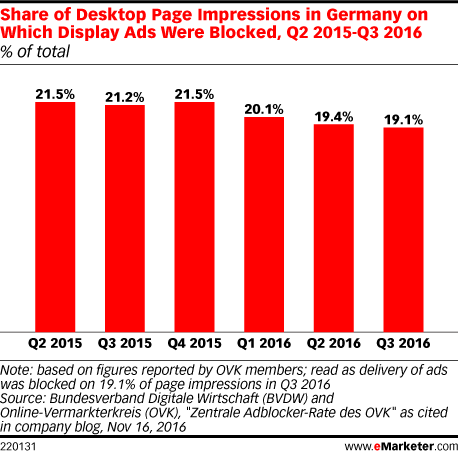}
\caption{Share of Desktop Page Impressions in Germany on Which Display Ads Were Blocked according to a report by BVDW and OVK \cite{declineGermany}}
\label{fig:declineGermany}
\end{figure}

On the other hand, Online-Vermarkterkreis (OVK) and Bundesverband Digitale Wirtschaft (BVDW) published a report titled “Zentrale Adblocker-Rate des OVK” \cite{declineGermany} in December 2016, that shows a decline in the "incidence rates" related to ad-blockers from the first quarter to the third quarter of 2016 (Figure \ref{fig:declineGermany}). 

The critical difference to be noted in the two sources cited above is that PageFair looked at the number of penetration or installations of Ad-blockers whereas BVDW \& OVK focused on the actual number of websites that were affected by such Ad-blockers.

\textbf{Working of Ad-blockers.} Ad-blocker tools work mostly on removing browser page elements. They look for certain HTML, DOM or CSS elements in the web page and process them to be removed. Ad-blockers usually work on a set of rules that indicate which such elements must be removed. These rules are part of "filter lists" such as EasyList \cite{easyList}. There are also user privacy protection lists such as EasyPrivacy \cite{easyPrivacy} which remove trackers. Another technique that Ad-blockers use is web request blocking, where these tools remove URLs that correspond to any publisher.

\begin{figure}
\centering
\includegraphics[height=2.8cm]{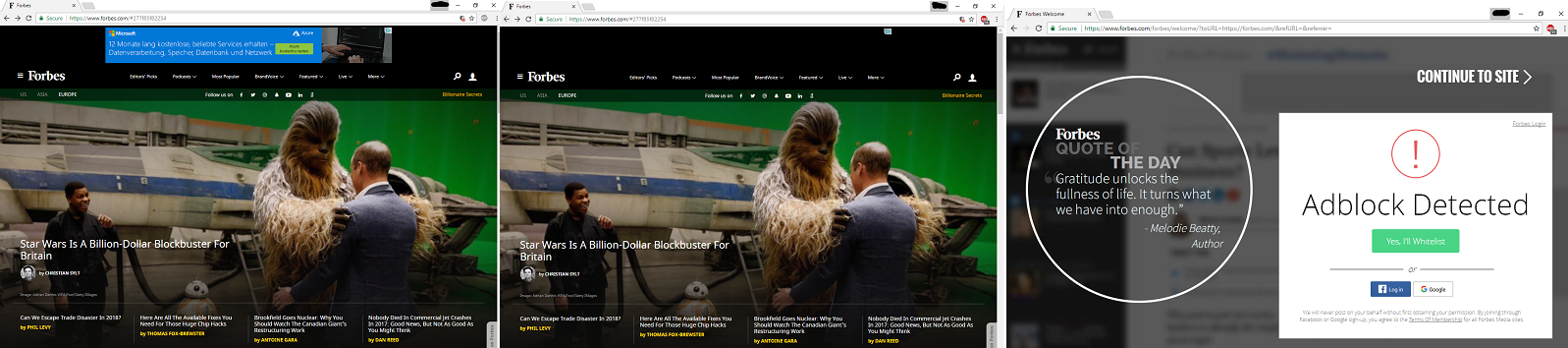}
\caption{Web page evolution for Forbes shows from left to right web page with ads, with Ad-blocker enabled and with an Anti Ad-blocker.}
Source: \url{www.forbes.com}
\label{fig:forbes}
\end{figure}

\textbf{How Anti Ad-blockers work.} Anti Ad-blocker scripts detect the presence of Ad-blockers and displays appropriate messages such as asking users to turn off Ad-blocker tools or not allowing users to view content. One such script has been published by IAB that "DEAL"s (Detect, Explain, Ask, Limit) with such Ad-blockers \cite{IAB2017}.

Anti Ad-blocker scripts are known to perform couple of operations. The first is to detect any Ad-block tool being used and the second is to notify the user to either disable Ad-block or to whitelist their website. These scripts may range from using first-party domains that check only aesthetic attributes such as height or width of ads to more complicated scripts from third-party domains that provide baits such as time delays, continuous detection or even use cookies to track Ad-block detection. It is a big challenge to detect such scripts since they are obfuscated and hidden deep into the system.

\begin{figure}
\centering
\includegraphics[height=4.5cm]{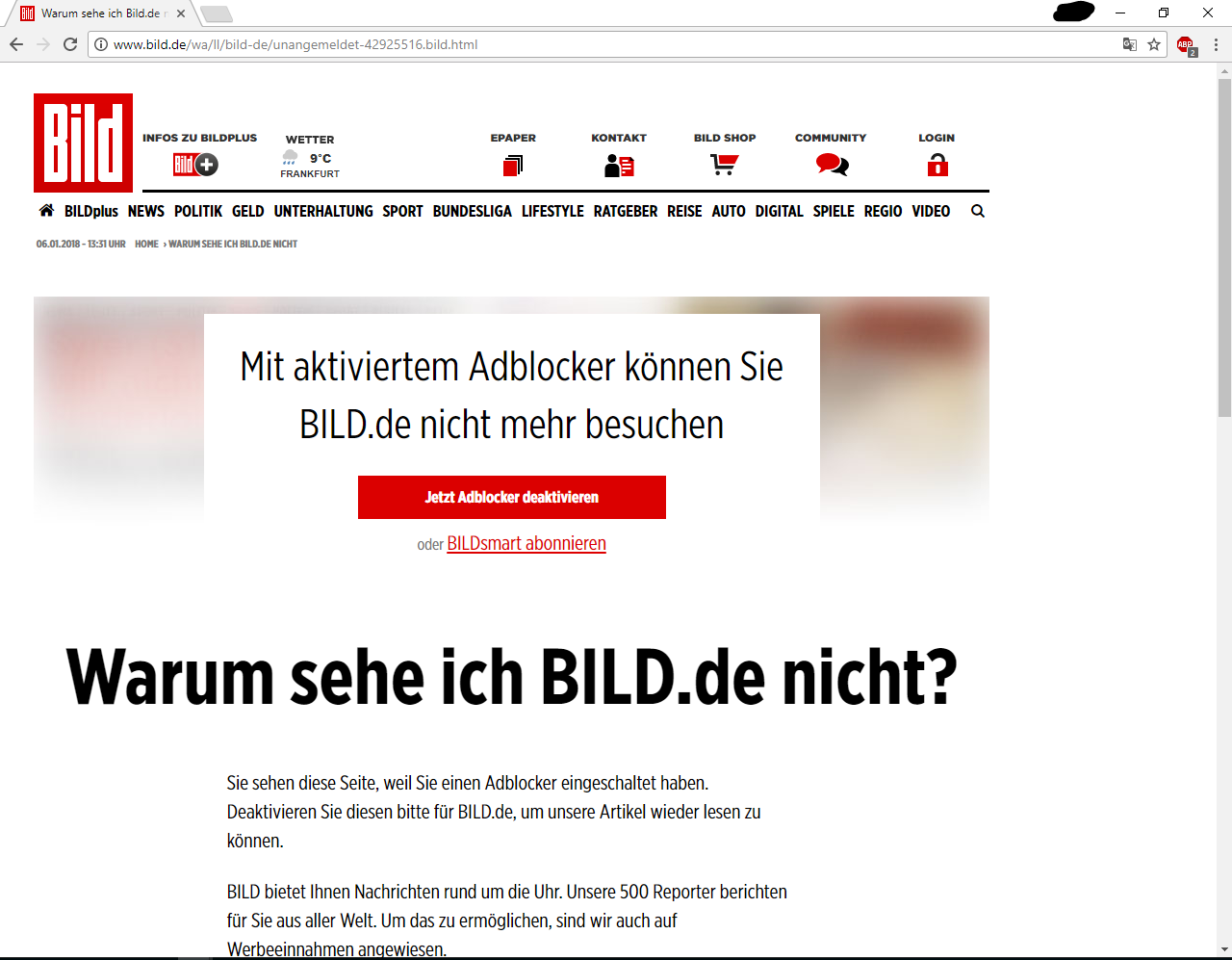}
\includegraphics[height=4.5cm]{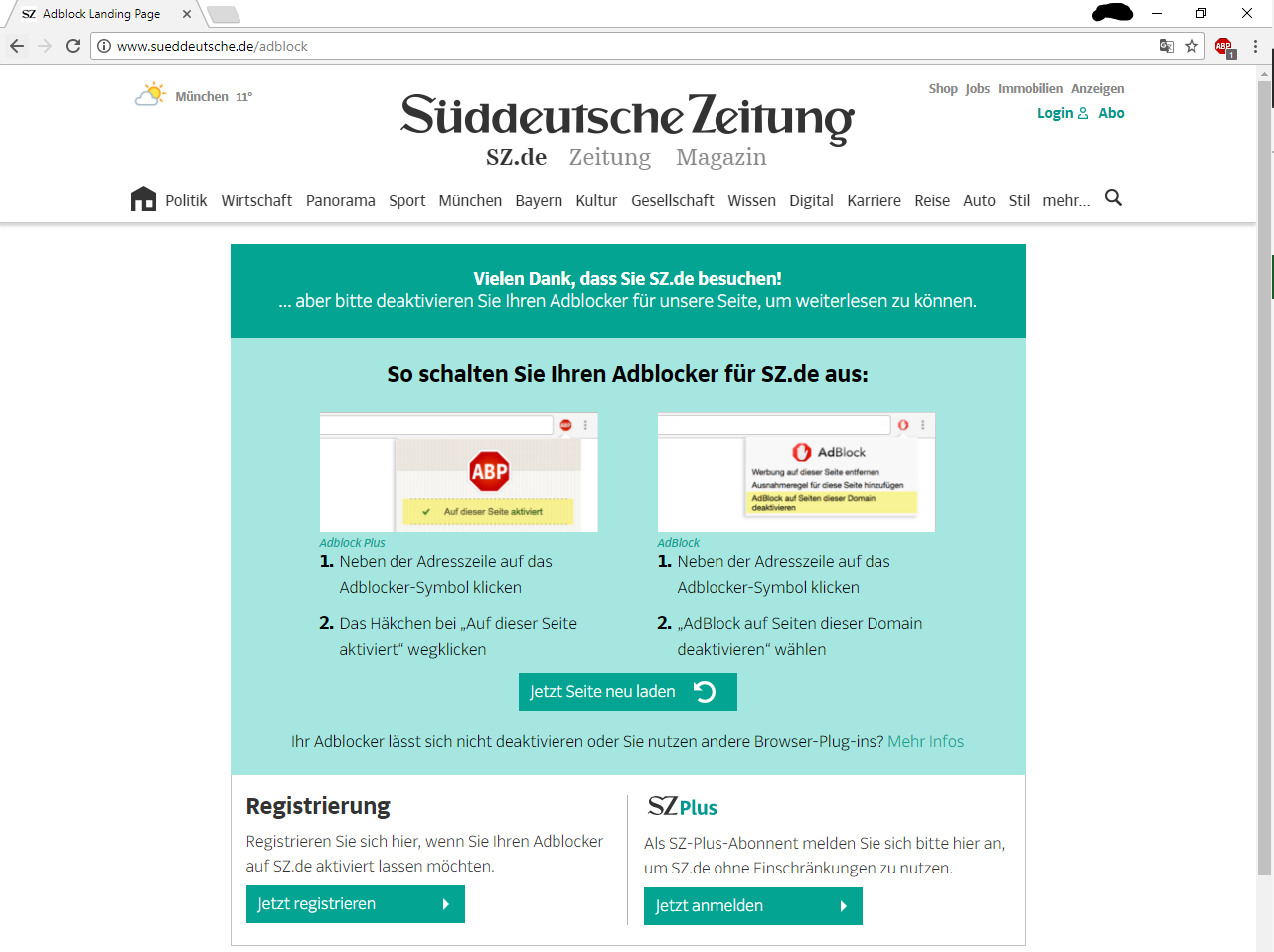}
\caption{German websites using Ad-blocker detection scripts.}
\label{fig:germanBlocks}
\end{figure}

Figure \ref{fig:germanBlocks} shows examples of German websites that show Ad-block detection responses. This is done by employing Anti Ad-blocking tools or scripts such as the one by IAB which detect the presence of such Ad-blockers.

\begin{figure}
\centering
\includegraphics[height=6.2cm]{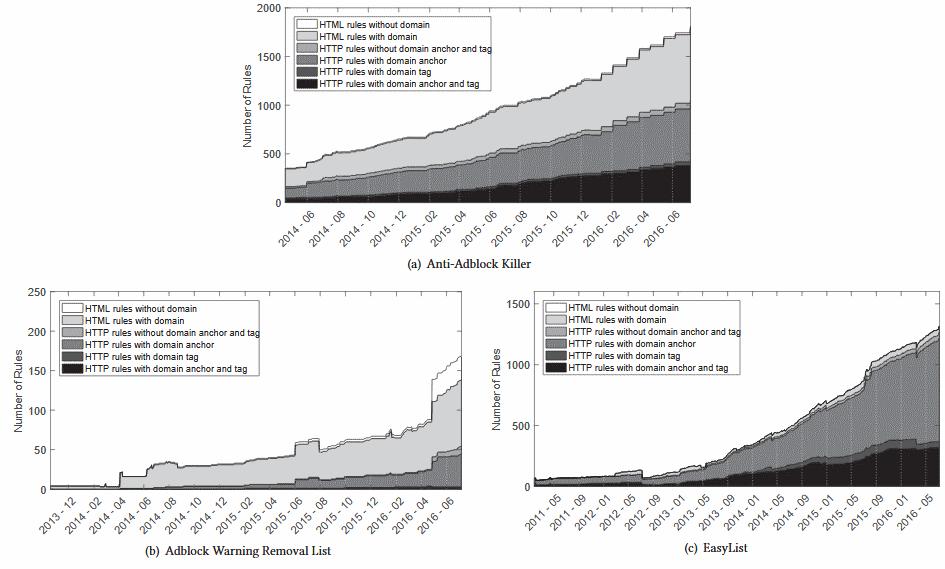}
\caption{Anti Ad-blocking scripts with their gradual rise in usage taken from the paper by Iqbal et. al. \cite{popAnti2018}}
\label{fig:anti-study}
\end{figure}

\textbf{Impact of Anti Ad-blockers.} "The Ad Wars: Retrospective Measurement and Analysis of Anti-Adblock Filter Lists" is a paper published jointly by scientists at University of Iowa and University of California-Riverside. In this paper, Iqbal et. al. scan Alexta Top 5K websites for Anti Ad-block tools such as Anti Ad-block Killer List \cite{aakl} and Combined EasyList for Anti Ad-block scripts. The results as shown in (Figure \ref{fig:anti-study}) taken from their paper, showed a gradual increase in Anti Ad-blocking scripts, which indicate that the online advertising industry reacting to the immediate losses.

\subsection{Related Work}
Our paper discusses in length the Anti Ad-blocking techniques and their effects adopted mainly by Nithyanand et al. \cite{Rishab2016} and Haris et al. \cite{Haris2015}. Nithyanand et al. \cite{Rishab2016} in their paper "Adblocking and Counter-Blocking: A Slice of the Arms Race", found that 6.7\% from the top 5K Alexa websites deploy Anti Ad-blockers. They also consider all websites that include Anti Ad-block scripts in their analysis. 

Haris et al. \cite{Haris2015} in their paper "Detecting Anti Ad-blockers in the Wild", noted that 686 websites in the Alexa top 100K use Anti Ad-blockers on their home page. They also notice different notifications ranging from asking users to disable Ad-blockers to paying a subscription fee.

Garimella et al. \cite{Garimella2017} in their paper "Ad-blocking: A Study on Performance, Privacy and Counter-measures", discuss extensively multiple Ad-Blocker tools and compare their performance and privacy facets. 

\section{Data Collection}
As part of our data collection, we collected top websites from Amazon Alexa Website Rankings: \textit{https://www.alexa.com/topsites}. Publicly available data is restricted to only 50 per category out of the 500 available ranked websites. The ranking is done based on the average daily visitors to a site and also over the pageviews for that site for the last month. We subscribed to a 7-day free trial to collect top websites for our different category use cases. Our data sets include top 500 news websites, top 500 websites in Germany and a unique subset of top websites in the DACH (500 each for Germany, Austria and Switzerland) region.

To pull this data, we had to generate an access key with an Amazon AWS account which grants policy rights to AWIS (Amazon Web Information Services) located here: \textit{https://aws.amazon.com/awis/}. Furthermore, we had to subscribe to the 7-day Amazon Alexa free trial so that we can pull the information from Alexa using API calls. Listing \ref{listing:canQuery} shows an HTTP request using query parameters used to pull data using this API. We used Java to pull this information. We made use of the sample code provided by Amazon AWIS (\textit{https://aws.amazon.com/awis/getting-started/}) and modified it to our needs to pull the site names.

\begin{listing}[!h]
\begin{minted}[mathescape,
               linenos,
               numbersep=5pt,
               frame=lines,
               baselinestretch=1.2,
			   fontsize=\footnotesize,
               escapeinside=||,
               framesep=2mm]{js}	
https://awis.amazonaws.com/api?Action=CategoryListings&Count=20
	&Descriptions=True&Path=Top%2FNews
	&Recursive=False&ResponseGroup=Listings
    &SortBy=Popularity&Start=1
\end{minted}
\caption{An HTTP request to AWIS to pull top news sites}
\label{listing:canQuery}
\end{listing}

A maximum of 20 results can be pulled at each HTTP request to the host \textit{"awis.amazonaws.com"} with a starting count of counter = 1 (Start). This counter is incremented by 20 for every request, till all the data has been pulled in. The code uses SHA-256 hashing algorithm to create an Authorization that includes the Access Key obtained earlier.

The response contains a list of websites sorted according to the parameters specified. A sample response is listed in Listing \ref{listing:httpResponse}. The response is then parsed using an XML parser to obtain each website name and is stored in a file. For generating the unique subset of all the websites in the DACH region, we created a HashSet that does not allow duplicates to be entered into the set. This is done so that common websites is only taken into consideration once.

\begin{listing}[!h]
\begin{minted}[mathescape,
               linenos,
               numbersep=5pt,
               frame=lines,
               baselinestretch=1.2,
			   fontsize=\footnotesize,
               escapeinside=||,
               framesep=2mm]{js}	
<aws:CategoryListingsResponse xmlns:aws="http://alexa.amazonaws.com/doc/2005-10-05/">
<aws:Response xmlns:aws="http://awis.amazonaws.com/doc/2005-07-11">
<aws:OperationRequest>
<aws:RequestId>3ffe5d1a-8939-5b7a-9002-bec5a1243i9a</aws:RequestId>
</aws:OperationRequest>
<aws:CategoryListingsResult>
<aws:Alexa>
  <aws:CategoryListings>
    <aws:RecursiveCount>21</aws:RecursiveCount>
    <aws:Count>2</aws:Count>
    <aws:Listings>
      <aws:Listing>
        <aws:DataUrl type="navigable">http://www.reddit.com</aws:DataUrl>
        <aws:Title>Reddit.com</aws:Title>
        <aws:PopularityRank>2</aws:PopularityRank>
      </aws:Listing>
      <aws:Listing>
        <aws:DataUrl type="navigable">http://www.cnn.com/</aws:DataUrl>
        <aws:Title>CNN</aws:Title>
        <aws:PopularityRank>3</aws:PopularityRank>
      </aws:Listing>
    </aws:Listings>
  </aws:CategoryListings>
</aws:Alexa>
</aws:CategoryListingsResult>
<aws:ResponseStatus xmlns:aws="http://alexa.amazonaws.com/doc/2005-10-05/">
<aws:StatusCode>Success</aws:StatusCode>
</aws:ResponseStatus>
</aws:Response>
</aws:CategoryListingsResponse>
\end{minted}
\caption{An HTTP response in XML for top news sites.}
\label{listing:httpResponse}
\end{listing}

These websites are then sent through our data processor described in the next section.

\section{Anti Ad-blockers Detection}

In this paper, we are trying to replicate and reproduce the work done by Haris et al. \cite{Haris2015} in their paper "Detecting Anti Ad-blockers in
the wild", and by Rishab et al. \cite{Rishab2016} in their paper "Adblocking and Counter-Blocking: A Slice of the Arms Race", but in a different context. Haris et al. used websites from the Alexa top-100K which show a slight bias towards the American Ad-blocking industry.   
This is evident from the country of origin of Alexa top-100 websites listed on Alexa's website \cite{Top-100} where Google Germany and Amazon Germany are the only German sites in that list.

Our focus in this paper are the popular sites in Germany, DACH region (Germany, Austria, Switzerland) and we also take a look at the sub-categories of News websites. We plan to compare our results with the results obtained by Haris et al. and Rishab et al. thereby producing an analogy to their findings. Our assumption is that our results will be similar to theirs.
The previously mentioned papers also do not provide any source code to their technique but explain their methodology in the paper which has been adopted by us. 
In this section, we try to explain how we have adopted and modified the methodology proposed by them.

\subsection{Overview}

As mentioned earlier in this paper, HTML and DOM elements in a browser page contains source code of websites which use Anti Ad-blockers and which do not. Websites rely mostly on adding and/or modifying HTML content for detecting Ad-blocker usage.

Haris et al. in their paper categorize these changes into the following categories:

\textit{Nodal Changes:} These changes are concerned with the addition of extra HTML DOM elements.

\textit{Style Changes:} These changes are concerned with the modification of HTML DOM elements relating to their style and appearance attributes.

\textit{Textual Changes:} These changes are concerned with the modification in textual elements in any HTML page.

\textit{Structural Changes:} These changes are concerned with the modification in elements such as the innerHTML DOM property and whether any changes lead to redirection to another URL.

For websites that show such changes as mentioned above with Ad-blockers enabled, they are said to be using Anti Ad-blockers.

\subsection{Methodology}
The methodology we use is to capture and record these changes made by Websites and create a model to predict the usage of Anti Ad-blocker scripts. To not over complicate the technique and simplify the process, we consider only Nodal Changes, Textual Changes and Structural Changes as described in the previous section and altogether ignore Style changes and the innerHTML component of Structural changes from our experiments.
Similar to Haris et al. we perform A/B testing using Selenium Chrome WebDriver \cite{selenium} using an extension of Ad-blocker in one instance and not using one in another instance.

\textbf{Browser instance automation}
To automate this process, we make use of Selenium Chrome Webdriver \cite{Chrome Webdriver} to launch two separate instances of Google Chrome web browser one with an Ad-block extension installed and one without one.
We open the sites under test with each instance.

\textbf{Browser profiles}
We create two new browser profiles in Google Chrome. In one of the profiles we install Ad-blocker extension and in the other we do not. This provides a clean system under test not subject to any previous website caching for that profile. We use Adblock Plus \cite{Google2018} due to its popularity. 
It has options to configure various filter lists such as \cite{filterLists}. We use the EasyList filter \cite{easyList} after removing all the Anti Ad-block rules as well as disallow any possible Acceptable Ads. The Anti Ad-block rules are marked by the comments ! Anti Ad-block in the filter list.

\textbf{Python Script for Data Collection}
A script in python was used for the data collection purpose.
It reads a list of websites and then checks if the website is live.
Then it opens the same website in two modes: One with Ad-block installed and one without and takes a screen grab (which can be used later to verify the results) and saves the HTML content for both cases.

BeautifulSoup library \cite{beautifulSoup} is used for parsing the HTML file and recording the difference of various features and the presence of certain keywords in the text such as adblocker, adblock, ad block, ad-block, whitelist, block-adblock, pagefair, etc. This is then written to a csv file.

\textit{Nodal Features:} We consider the following nodal features for our processing: anchor, div, h1, h2, h3, img, table, p, iframe and the text nodes.

\textit{Textual features:} We consider the number of lines, words and the text characters.The following key words are also used in method A of the A/B testing: adblocker, adblock, ad block, ad-block, whitelist, block-adblock, pagefair, fuckadblock.

\textit{Structural features:} We note any URL redirections by using a simple yes or a no in generation of the csv file.

\textbf{Creating and Training the model}
For the creation and training of the model, we used the Weka toolkit \cite{weka} that provides out of the box machine learning algorithms used for classifications. It also has a Java interface and a GUI which makes it easier to work with the data.

We used the following machine learning classifiers such as J48 Decision Tree, Random Forest and Naive Bayes which can then be used to identify websites employing anti Ad-blockers. 

For the learning phase, we required labeled data. 
No such public database is available so it need to be done manually.
For this purpose, we need some negative samples (websites that do not employ Anti Ad-blockers) and positive samples (websites that employ Anti Ad-blockers). 

To collect positive samples we look into the issue list of anti-adblock-killer \cite{Reek issues}, it is used for blocking anti-adblockers so naturally websites on its issue list use Anti ad-blockers.
We scraped the websites list from its issue list and manually checked them to confirm this. This formed our positive samples.

For negative samples, we collected a list of global websites using a collection of sources such as Amazon Alexa Web Ranking, Similarweb \cite{similarweb} and Quantcast \cite{qc} and manually checked them. We then ran these websites through our Python data collection script and manually labelled the websites in the csv file.

This was used as our training set and we ran this through various Weka classifiers to create our model.

\textbf{Analysis of the features}
Our next step is to perform feature analysis on the features that we had generated in the previous section from the training data.
Different attributes have different weights for discriminating between the classes to be learned. 
We use information gain ratio \cite{IG} for the same.

\begin{table}[]
\centering
\caption{Features ranked based on Information Gain}
\begin{tabular}{@{}ll@{}}
\toprule
Information Gain & Feature    \\ \midrule
25.65\%          & lines      \\
20.66\%          & p          \\
20.60\%          & a          \\
20.44\%          & div        \\
19.2\%           & words      \\
18.05\%          & tags       \\
16.12\%          & img        \\
11.76\%          & h1         \\
11.05\%          & keyword    \\
11.05\%          & h3         \\
9.15\%           & iframe     \\
8.67\%           & h2         \\
4.87\%           & table      \\
0.78\%           & url change \\ \bottomrule
\end{tabular}
\label{table:IG}
\end{table}

The rankings shown in Table \ref{table:IG} show that textual features have higher information gain which is similar to the results
of Haris et al. although the information gain from the attributes maybe different.

\subsection{Evaluating Classifier Models}
Building on the previous work we perform different machine learning classification methods using the Weka toolkit. We use the most common ROC metrics such as precision, recall, and the area under ROC curve (AUC) to compare the different classifier algorithms used in our experiments.

The following tables from \ref{table:RF acc} to \ref{table:J48 cm} summarizes the classification accuracy and the effectiveness of the classifiers we use for our training model.

\begin{table}[]
\centering
\caption{Random Forest - Detailed Accuracy By Class}
\label{table:RF acc}
\begin{tabular}{@{}lllllllllll@{}}
\toprule
              & TP Rate & FP Rate & Precision & Recall & F-Measure & MCC   & ROC Area & PRC Area & Class &  \\ \midrule
              & 1.000   & 0.024   & 0.993     & 1.000  & 0.996     & 0.984 & 0.999    & 1.000    & FALSE &  \\
              & 0.976   & 0.000   & 1.000     & 0.976  & 0.988     & 0.984 & 0.999    & 0.998    & TRUE  &  \\
Weighted Avg. & 0.995   & 0.019     & 0.995  & 0.995     & 0.994 & 0.984    & 0.999    & 0.999 &  \\ \bottomrule
\end{tabular}
\end{table}

\begin{table}[]
\centering
\caption{Random Forest - Confusion Matrix}
\label{table:RF cm}
\begin{tabular}{@{}lll@{}}
\toprule
a   & b   & \textless-- classified as \\ \midrule
422 & 0   & a = FALSE                 \\
3   & 121 & b = TRUE                  \\ \bottomrule
\end{tabular}
\end{table}

\begin{table}[]
\centering
\caption{Naive Bayes - Detailed Accuracy By Class}
\label{table:NB acc}
\begin{tabular}{@{}llllllllll@{}}
\toprule
              & TP Rate & FP Rate & Precision & Recall & F-Measure & MCC   & ROC Area & PRC Area & Class \\ \midrule
              & 0.976   & 0.798   & 0.806     & 0.976  & 0.883     & 0.304 & 0.817    & 0.931    & FALSE \\
              & 0.202   & 0.024   & 0.714     & 0.202  & 0.314     & 0.304 & 0.816    & 0.545    & TRUE  \\
Weighted Avg. & 0.800   & 0.622   & 0.785     & 0.800  & 0.754     & 0.304 & 0.817    & 0.844    &       \\ \bottomrule
\end{tabular}
\end{table}

\begin{table}[]
\centering
\caption{Naive Bayes - Confusion Matrix}
\label{table:NB cm}
\begin{tabular}{@{}lll@{}}
\toprule
a   & b  & \textless-- classified as \\ \midrule
412 & 10 & a = FALSE                 \\
99  & 25 & b = TRUE                  \\ \bottomrule
\end{tabular}
\end{table}

\begin{table}[]
\centering
\caption{J48 Decision Tree - Detailed Accuracy By Class}
\label{table:J48 acc}
\begin{tabular}{@{}llllllllll@{}}
\toprule
              & TP Rate & FP Rate & Precision & Recall & F-Measure & MCC   & ROC Area & PRC Area & Class \\ \midrule
              & 0.988   & 0.250   & 0.931     & 0.988  & 0.959     & 0.806 & 0.896    & 0.943    & FALSE \\
              & 0.750   & 0.012   & 0.949     & 0.750  & 0.838     & 0.806 & 0.896    & 0.837    & TRUE  \\
Weighted Avg. & 0.934   & 0.196   & 0.935     & 0.934  & 0.931     & 0.806 & 0.896    & 0.919    &       \\ \bottomrule
\end{tabular}
\end{table}

\begin{table}[]
\centering
\caption{J48 Decision Tree - Confusion Matrix}
\label{table:J48 cm}
\begin{tabular}{@{}lll@{}}
\toprule
a   & b  & \textless-- classified as \\ \midrule
417 & 5  & a = FALSE                 \\
31  & 93 & b = TRUE                  \\ \bottomrule
\end{tabular}
\end{table}

\section{Results Analysis}
\subsection{Analysis of Test Set}

Once we have trained our models, we need to run our test set which is the website list that we collected earlier through the Weka toolkit to generate the classification metrics. For each region and category, we generate the True Positives (TP), False Positives (FP) and the Precision involved for each of the three classifiers being used. We do not consider the True Negatives (TN) and the False Negatives (FN) due to the manual work and limitations for the time frame for this paper.

\textbf{Germany Region.}
We collected 500 top websites belonging to Germany region from Alexa Website Rankings \textit{https://www.alexa.com/topsites} for our data analysis. Running it through our methodology we were able to successfully identify and create a test set of 418 websites. The results are listed in table \ref{table:DE}.

\begin{table}[]
\centering
\caption{TP, FP, Precision of Classifiers in Germany region}
\label{table:DE}
\begin{tabular}{@{}llll@{}}
\toprule
& Naive Bayes Classifier & J48 Classifier & Random Forest Classifier       \\ \midrule
Predicted              & 4             & 24                       & 26\\
TP                     & 3              & 8                       & 8\\
FP                     & 1              & 16                       & 18\\
Precision              & 0.75          & 0.333                    & 0.308\\ \bottomrule
\end{tabular}

\end{table}

We observe that 8 websites from the 418 websites deploy an Anti Ad-blocker of some sorts. This accounts to 1.9\% of the websites we identified for Germany region.

\textbf{DACH Region.}
We collected 500 top websites belonging to Germany, Austria and Switzerland (DACH region) from Alexa Wesbite Rankings 

\textit{https://www.alexa.com/topsites} for our data analysis. From this set of 1500 websites, we took a unique set of websites, removing any duplicate websites that exist for each country. This accumulated to a total of 809 unique websites. The results for this set are listed in table \ref{table:DACH}. 

\begin{table}[]
\centering
\caption{TP, FP, Precision of Classifiers in DACH region}
\label{table:DACH}
\begin{tabular}{@{}llll@{}}
\toprule
& Naive Bayes Classifier & J48 Classifier & Random Forest Classifier       \\ \midrule
Predicted              & 10             & 40                       & 41\\
TP                     & 3              & 11                       & 10\\
FP                     & 7              & 29                       & 31\\
Precision              & 0.3          & 0.275                    & 0.244\\ \bottomrule
\end{tabular}

\end{table}

We observe that 11 out of the 809 websites deploy an Anti Ad-blocker of some sorts. This accounts to 1.4\% of the websites we identified for DACH region.

\textbf{News Category.}

We collected 500 top websites belonging to News Category from Alexa Website Rankings \textit{https://www.alexa.com/topsites} for our data analysis. Running it through our methodology we were able to successfully identify and create a test set of 357 websites. The results are listed in table \ref{table:News}.

\begin{table}[]
\centering
\caption{TP, FP, Precision of Classifiers in News Category}
\label{table:News}
\begin{tabular}{@{}llll@{}}
\toprule
& Naive Bayes Classifier & J48 Classifier & Random Forest Classifier       \\ \midrule
Predicted              & 7             & 18                       & 16\\
TP                     & 6              & 10                       & 11\\
FP                     & 1              & 8                       & 5\\
Precision              & 0.857          & 0.556                    & 0.688\\ \bottomrule
\end{tabular}

\end{table}

We observe that 11 out of the 357 websites deploy an Anti Ad-blocker of some sorts. This accounts to 3.1\% of the websites we identified for News category.

\textbf{Comparison.}
We observed during verification that some websites have certainly designed different methods to evade any kind of Ad-blocker detection. This includes, but is not limited to ads that are allowed under the Acceptable Ads programme \cite{accads}. However we do not consider such a behaviour in generating the analysis metrics.

Haris et al. \cite{Haris2015} observed in their paper that out of the Alexa top 100K websites they evaluated, they were able to identify 686 websites that used Anti Ad-blockers using their methodology.

\subsection{Types of Ad-block Detection Responses}
We manually collected and verified the positive samples for Ad-block detector responses. We use this to categorize how they behave when encountering an Ad-blocker.

\begin{figure}
\centering
\includegraphics[height=5.8cm]{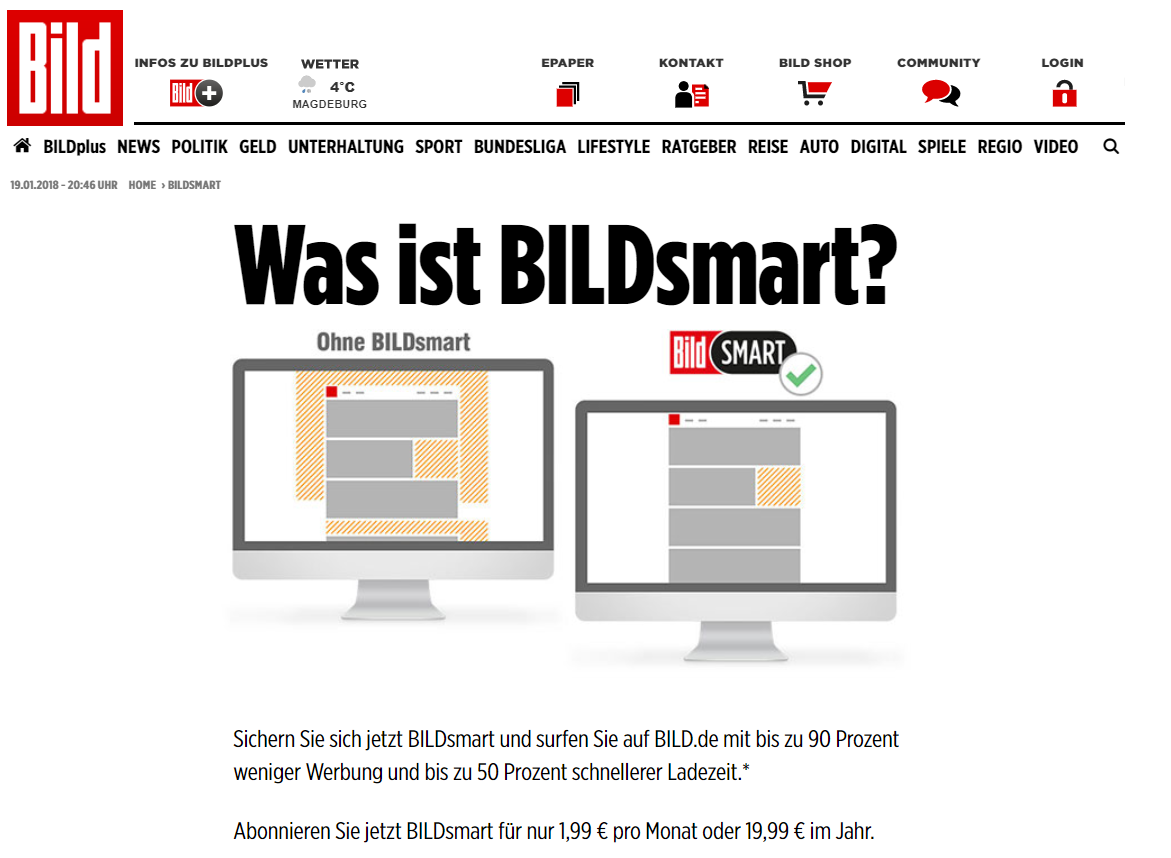}
\caption{Bild.de ask users to pay to view up to 90 \% less adverts.}
\label{fig:bildde-pro}
\end{figure}

\begin{figure}
\centering
\includegraphics[height=7.1cm]{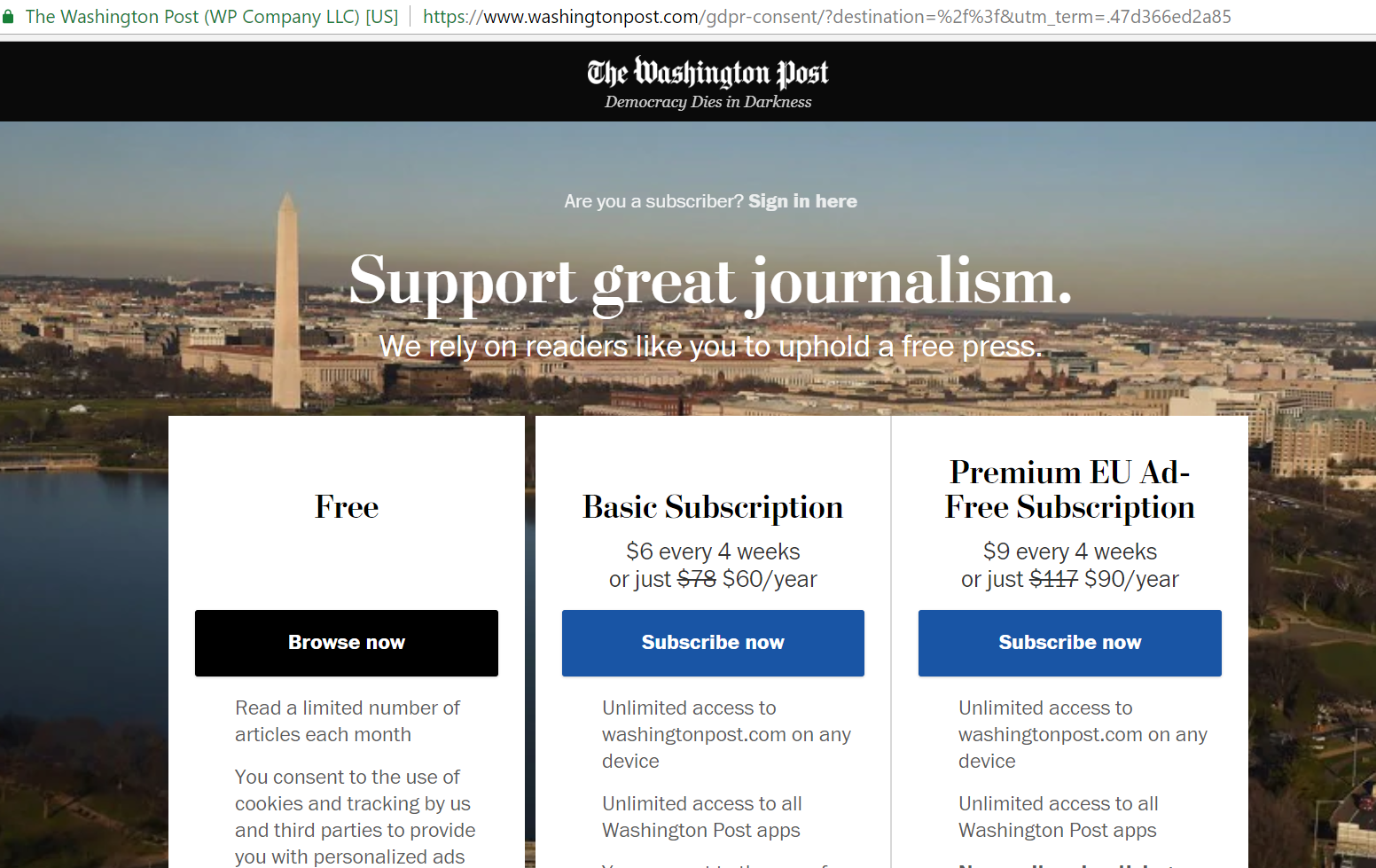}
\caption{The Washington Post asks users to pay to view content based on region.}
\label{fig:twp1}
\end{figure}

We also measure the extent to which websites go to inform or request users from disabling Ad-blockers. We define a new terminology for this measure called "CIA". It involves the following three types:

\textbf{Cost Model (C):} We define this type as a cost or monetization model. This includes websites that ask for paid subscriptions (Figure \ref{fig:bildde-pro}), content for a limited number of days (Figure \ref{fig:twp1}) or even donations.

\begin{figure}
\centering
\includegraphics[height=5.4cm]{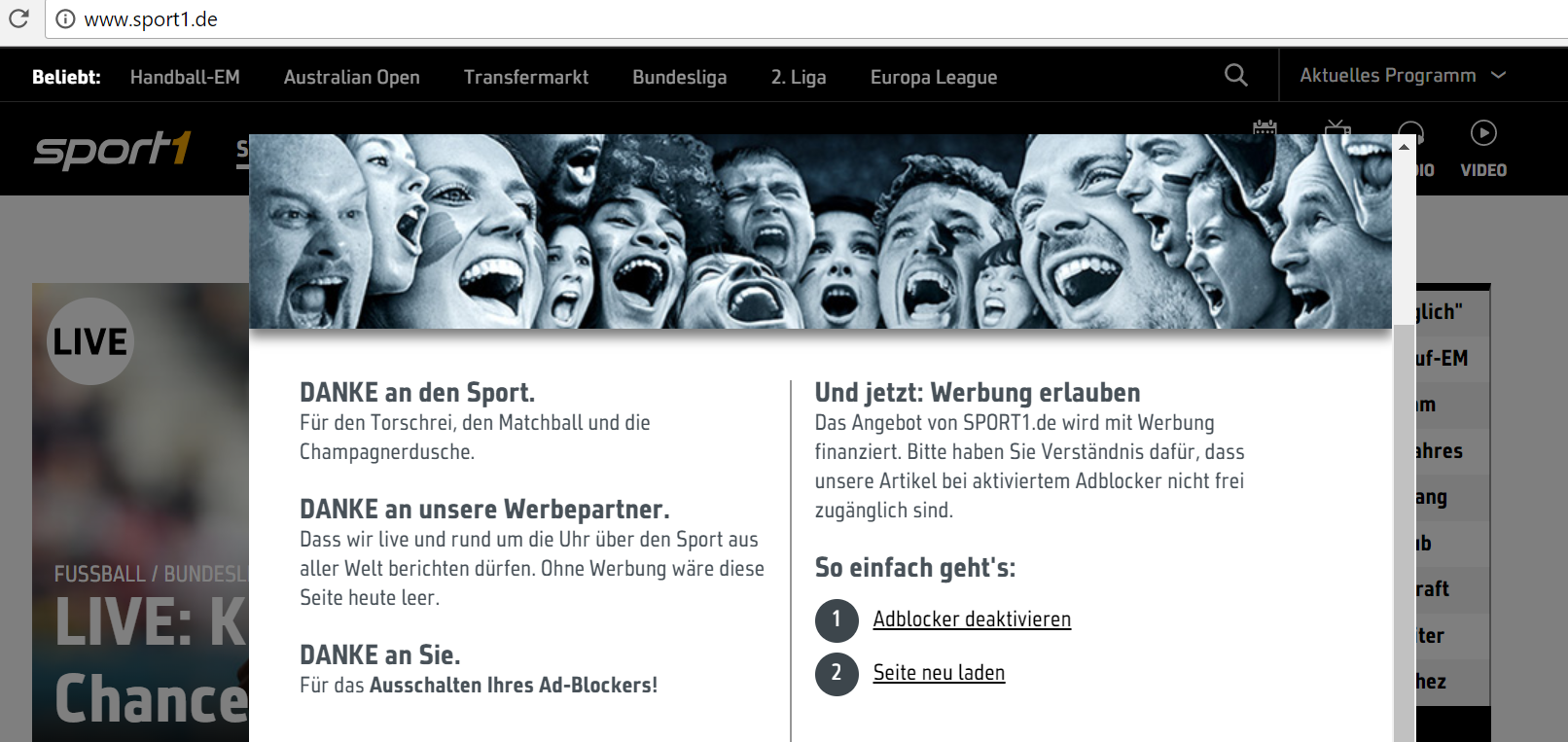}
\caption{Sport1.de do not allow content to be viewed until Ad-blocker is disabled.}
\label{fig:sport1}
\end{figure}

\textbf{Invisibility (I):} We define this type as not allowing users to view content making their content invisible until Ad-blockers are disabled. Two stand out examples of this measure are: \url{www.bild.de} or \url{www.sport1.de} (Figure \ref{fig:sport1}). 

\begin{figure}
\centering
\includegraphics[height=5.5cm]{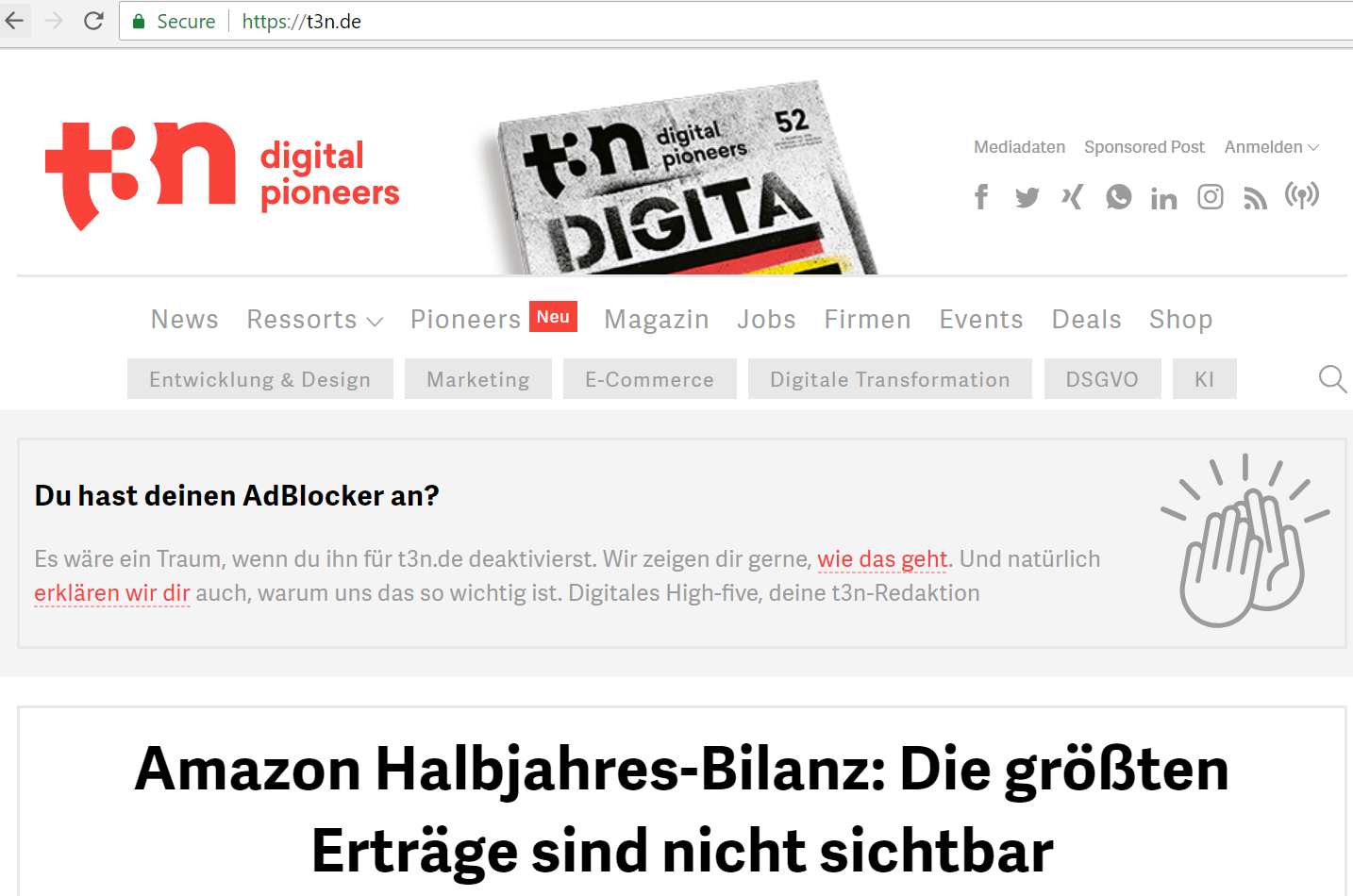}
\caption{t3n.de asks users to remove Ad-blockers but do not block content.}
\label{fig:t3n}
\end{figure}

\textbf{Availability (A):} We define this type as a conservative approach to dealing with Ad-blockers. It is demonstrated by \textit{https://t3n.de/} as shown in (Figure \ref{fig:t3n}). They still allow users to view content on their website but ask them politely to disable Ad-blockers, thereby enabling availability of content.

\subsection{Geographical Comparison}
In their paper, Haris et al. \cite{Haris2015} identified 686 websites from the Alexa top 100K websites that deploy Anti Ad-blockers. This list mostly details websites concerning the United States of America or China. Only a few websites from this list such as \url{www.google.de} and \url{www.amazon.de} belong to the Germany region.

Rishabh et al. \cite{Rishab2016} in their paper note that 6.7\% of websites in a specific category such as News deploy Anti Ad-Blocker.

In our analysis, we observe that 1.9\% of the 418 identified websites in Germany region employ Anti Ad-blockers. For DACH region, this percentage stands at 1.4\% for 809 websites. For News category, this percentage changes to 3.1\% out of the 357 websites. Our methodology contains various limitations which is described in the next section.

\subsection{Limitations}

Our methodology cannot detect all websites that employ Ad-block detection due to some limitations. They are described below:
\begin{itemize}
\item{We remove all anti Ad-block filters in EasyList. This would probably not be the default settings used by users across the goal. But as we are attempting to quantify websites which employ Anti Ad-blockers, we feel that this configuration will give us a clearer picture.}
\item{We only check anti Ad-blockers on the home page and not on any linked or a sub page.}
\item{We look at websites that detect Ad-blockers and make any HTML or DOM changes to the browser page as using Anti Ad-blockers.}
\item{We use Adblock Plus for our measurements but there are other Ad-block tools which is not part of our study.}
\item{We also don't consider style features and cosine similarity features used by Haris et al. in their paper \cite{Haris2015}.}
\end{itemize}

\section{Impact of Anti Ad-blockers}
\subsection{Economical Impact}
\begin{figure}
\centering
\includegraphics[height=6.2cm]{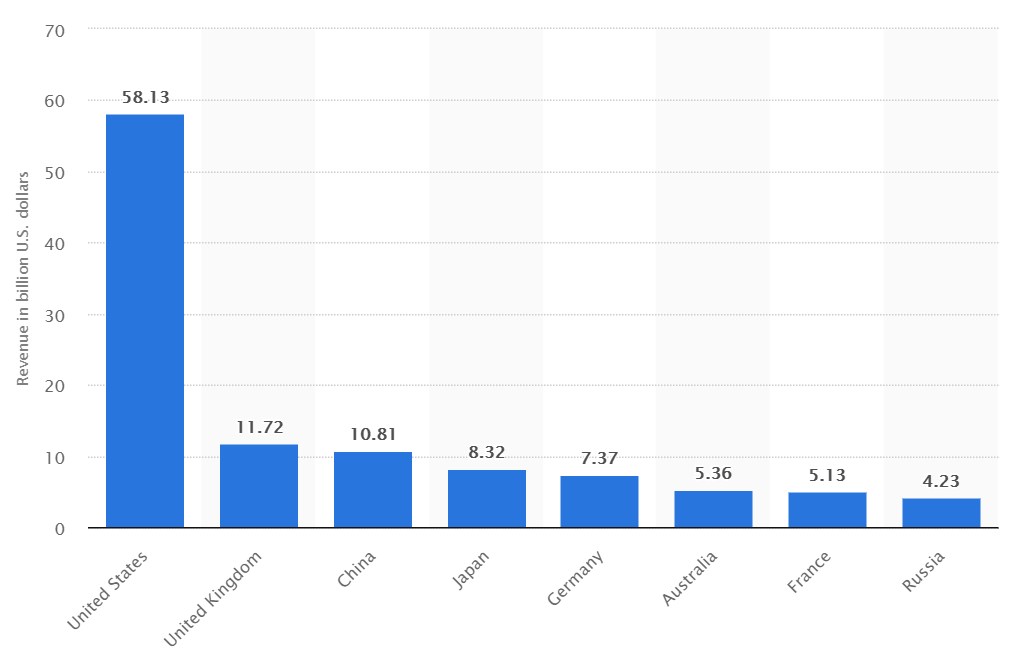}
\caption{Largest online advertisement markets in 2017 as published in the The Statistics Portal report titled "Online advertising revenue". \cite{largest2017}}
\label{fig:largestOnline2017}
\end{figure}

\textbf{Revenue Generation for Ads.} In a 2015 report titled "Digital advertising in Europe - Statistics \& Facts", The Statistics Portal \cite{europe2017} cited "online advertising revenues worldwide amounted to about \$170B, a figure that is expected to grow to more than \$330B by 2021. In a 2017 global comparison, the United Kingdom, Germany and France ranked among the largest online advertising markets in the world (Figure \ref{fig:largestOnline2017}), with digital ad revenues of \$11.72B, \$7.37B and \$5.13B, respectively. In Germany, the 2017 revenues stood at \$7.37B."

\subsection{Legality Aspects}
\textbf{European Union.} Appendix \ref{ePrivacyAppendix} details the Article 5.3 of the European Union's ePrivacy Directive. Under this EU law and the recently introduced GDPR regulation (described in a later section), publishers must ask for permission before accessing a user's personal information, similar to how websites must ask for permission to store cookies on user devices. But as publishers are only detecting ads delivery via HTML or DOM elements, they are complying with the ePrivacy directive.

\textbf{Germany.} Media groups Süddeutsche Zeitung, Pro-Sieben-Sat.1, and IP Deutschland, an RTL subsidiary recently fought a legal battle against Cologne-based Ad-blocking company Eyeo to ban Eyeo's Ad-blocking software Adblock Plus because of their revenue losses. According to a report by Wired.de \cite{wiredLegal} titled "Adblock Plus: Werbeblocker bleiben in Deutschland legal", Munich higher regional court ruled that the Ad-blocking software Adblock Plus is legal. The court also quoted that Eyeo is not dealing in "forbidden aggressive advertising" and that is why their software does not breach German laws. Although joining Eyeo's Adblock Plus whitelist is free, a few media groups are still filing cases against Eyeo because they are objecting that Eyeo is monetizing from advertisers joining these whitelists under the Acceptable Ads programme.

This court order comes is contrasting to another in Cologne. Digiday's article \cite{springerLoss} titled "What Axel Springer’s loss in ad-blocking suit means for UK publishers", describes the court judgment which states Adblock Plus must add Axel Springer to their Whitelist for free, but work within the confinements of the Acceptable Ads programme.

In a Hamburg judgement for Spiegel versus Eyeo, Eyeo cited \cite{hamburgJudgement} in the judgement "Urteil des Landgerichts Hamburg (Az. 315 O 293/15)", that as of August 2015: "Adblock Plus was installed on approximately 9.55 million browsers with German IP addresses, which accounted for around 5\% of the computers in Germany used to access the Internet". The judges finally concluded that users do not wish to see undesired advertising and they also want safety from any malicious softwares that might take control of their data.

In a blog post, "Adblock Plus and (a little) more", detailing the court decisions \cite{adblockLegal} Adblock Plus is currently still legal in Germany and as such the judges have not favored using Ad-blocking detectors. 

\subsection{Ethical Aspects}
The argument of Anti Ad-blocking ensues on the topic of maintenance of websites and henceforth businesses. Users of Ad-block tools are unaware as to how websites keep their business running. For most of such "free" sites, their revenue is generated through advertisements. Few websites use subscription services while others have non Internet based revenues model.

Whatever be the case, labor must be paid. It must be in accordance to what a writer, producer, musician, developer or any professional writes, creates or develops. Websites are paid according to impressions which are measured by metrics such as pay per click or cost per click on the advertisements. Small scale business are most likely to suffer because of their small workforce. Also, with increasing costs of living worldwide, publishers rely heavily on such revenue sources.

The trend among publishers is increasing with respect to the detection of Ad-blockers. They are asking users to either completely disable Ad-blockers or to pay money and subscribe to an ad-free version. This poses an ethical dilemma to users, on one hand to use an Ad-block and stopping revenue income for publishers and on the other hand for publishers, to use an Anti Ad-blocker to prevent users from blocking ads.

\subsection{Impact of GDPR}
The General Data Protection Regulation (GDPR) came into effect on the 25th of May, 2018 after a two year transitioning period in the European Union. The regulation carries a important changes to how user's personal data must be handled by companies. The GDPR does not allow companies to store or use user's personal data unless explicitly agreed by the user. Cookies, for example, are also considered to be part of this data. Users also have the right at any time to ask companies to delete their complete data.

Section 26 of the GDPR document is cited in Appendix \ref{GDPRAppendix}. As we have analysed in our discussion so far, detecting Ad-blockers work mostly on the concept of identifying HTML and DOM elements which are completely unrelated to the users. These HTML elements cannot be identified to any user and there is no such transmission of user’s personal data to any server, since it relies on client browser elements, which would lead to a non-compliance of the GDPR.

Many websites such as \url{npr.org} are providing only text content or country redirecting to an ad free EU version when users do not agree to provide access to their data. 

Of course, with advances in Ad-blocking detection techniques such as user targeted ads by collecting cookie information, there may be a case where a technical implementation could ideally violate the regulations of GDPR, but that is not in the scope of this paper.

\section{Alternatives to Anti Ad-blocking}
\subsection{Acceptable Ads Programme}
Acceptable ads programme (by Adblock Plus) \cite{accads} lays down a technique that is focused on effecting ads on websites. They define a certain set of criteria or rules that ensure that ads placed on websites are not annoying to users, disrupt or distort any of the primary web page content and are transparent such that there are no popups or any unseen advertisements that could potentially lead to nonacceptance of an advertisement on a website.

\subsection{Whitelists}
Websites that comply to the Acceptable Ads programme can get their websites "whitelisted" with Adblock Plus. This is very similar to anti-viruses marking some applications as "safe" or "non-malware" and allowing them to run, while blocking other potentially harmful ones. Publishers who employ Anti Ad-blockers ask users to whitelist their websites upon detection of an Ad-blocker. Users can by themselves Whitelist most websites but they usually do not change the default configuration.

\subsection{The rise of Anti Anti Ad-blockers}
Anti Ad-block killers are on the rise. They deploy tricks into thinking that the user is not using an Ad-blocker. The Ad-blocker blocker lets the user keep the Ad-blocker on, making everything look normal.

One such Anti Ad-block killer is the AAK: \textit{https://github.com/reek/anti-adblock-killer} \cite{AAK}. It is a JavaScript script with a default filter list similar to that of AdBlock Plus. The AakList filter list can also be configured on various Ad-block plugins.

\section{Conclusion and Future Work}
Our analysis involves quantitative and qualitative analysis, which builds on the work presented by Rishabh et al. \cite{Rishab2016} and Haris et al. \cite{Haris2015}. Our focus is mostly in the Germany and DACH region and a specific category which is News.

In our findings, we observed a range of 1.4-1.9\% of websites deploy Anti Ad-Blockers when it comes to a country or a specific region such as DACH and 3.1\% when it came to a specific category such as News. Rishabh et al. \cite{Rishab2016} in their paper found 6.7\% of websites in a category such as News deploy Anti Ad-Blocker. We also noted that Bayes Naive classifier performs better in our analysis than the analysis presented by Haris et Al. \cite{Haris2015}.

Our methodology depends on HTML and DOM element changes, which introduces a number of limitations. We discuss different characterizations of Anti Ad-Blocker detection responses and provide an economical and ethical overview. We also discuss on the legal aspects of Anti Ad-Blocker which is still legal according to the latest rulings in the court.

Our work concludes with looking at how GDPR impacts the Anti Ad-Blocker industry and we also provide insight into some alternatives to Anti Ad-Blocking which includes Acceptable Ads Programme and Whitelists.

\subsection{Future Work}
Future work for this area would include conducting a more exhaustive evaluation of websites in Germany and also refining our methodology to account for the fast-changing world of Ad-blockers and Anti Ad-blockers.

\subsection{Acknowledgements}
The authors would like to thank Professor Jens Grossklags for providing valuable inputs and insights.

\subsection{Source code and data release}
Source code for this project can be found at the link: \url{https://github.com/RohitPanda/seminar-blocking-adblockers}

\begin{subappendices}
\renewcommand{\thesection}{\Alph{section}}

\section{ePrivacy Directive 2002/58/EC of the European Parliament and of the Council}
Article 5.3 of the European Union's ePrivacy Directive \cite{eprivacy} states: "Member States shall ensure that the use of electronic communications networks to store information or to gain access to information stored in the terminal equipment of a subscriber or user is only allowed on condition that the subscriber or user concerned is provided with clear and comprehensive information in accordance with Directive 95/46/EC, inter alia about the purposes of the processing, and is offered the right to refuse such processing by the data controller. This shall not prevent any technical storage or access for the sole purpose of carrying out or facilitating the transmission of a communication over an electronic communications network, or as strictly necessary in order to provide an information society service explicitly requested by the subscriber or user."
\label{ePrivacyAppendix}

\section{GDPR Regulation (EU) 2016/679 of the European Parliament and of the Council}
Section 26 of the GDPR document \cite{GDPRtext} states, "The principles of data protection should therefore not apply to anonymous information, namely information which does not relate to an identified or identifiable natural person or to personal data rendered anonymous in such a manner that the data subject is not or no longer identifiable. This Regulation does not therefore concern the processing of such anonymous information, including for statistical or research purposes."
\label{GDPRAppendix}
\end{subappendices}

\end{document}